\documentclass{emulateapj}

\journalinfo{To appear in the Astrophysical Journal\\
{\it submitted 22 July 2006; accepted 7 February 2007}}
\submitted{To appear in the Astrophysical Journal}

\shorttitle{Six TrES Candidate Transiting Planets}
\shortauthors{O'Donovan et al.}

\newcommand{\tOne}{\mbox{T--And0--00948}} 
\newcommand{\tTwo}{\mbox{T--And0--01241}} 
\newcommand{\tThree}{\mbox{T--And0--02022}}
\newcommand{\tFour}{\mbox{T--And0--02462}} 
\newcommand{\tFive}{\mbox{T--And0--03874}}
\newcommand{\tSix}{\mbox{T--And0--03912}}

\newcommand{\first}{\mbox{HD\,209458\,b}}

\newcommand{\tresOne}{\mbox{TrES--1}}
\newcommand{\tresTwo}{\mbox{TrES--2}}

\newcommand\lsim{\mathrel{\rlap{\lower4pt\hbox{\hskip1pt$\sim$}}\raise1pt\hbox{$<$}}}
\newcommand\gsim{\mathrel{\rlap{\lower4pt\hbox{\hskip1pt$\sim$}}\raise1pt\hbox{$>$}}}

\begin{document}

\title{Outcome of Six Candidate Transiting Planets \\ from a T\lowercase{r}ES Field in Andromeda} 

\author{Francis~T.~O'Donovan\altaffilmark{1}, David~Charbonneau\altaffilmark{2,3}, Roi~Alonso\altaffilmark{4}, Timothy~M.~Brown\altaffilmark{5}, Georgi~Mandushev\altaffilmark{6}, Edward~W.~Dunham\altaffilmark{6}, David~W.~Latham\altaffilmark{2}, Robert~P.~Stefanik\altaffilmark{2}, Guillermo~Torres\altaffilmark{2}, Mark~E.~Everett\altaffilmark{7}}

\altaffiltext{1}{California Institute of Technology, 1200 East
  California Boulevard, Pasadena, CA 91125; ftod@caltech.edu}

\altaffiltext{2}{Harvard--Smithsonian Center for Astrophysics, 60
  Garden Street, Cambridge, MA 02138}

\altaffiltext{3}{Alfred P. Sloan Research Fellow}

\altaffiltext{4}{Laboratoire d'Astrophysique de Marseille, Traverse du Siphon, 13376 Marseille 12, France}

\altaffiltext{5}{Las Cumbres Observatory Global Telescope, 6720 Cortona Drive, Suite 102, Goleta, CA 93117}

\altaffiltext{6}{Lowell Observatory, 1400 West Mars Hill Road,
  Flagstaff, AZ 86001}

\altaffiltext{7}{Planetary Science Institute, 1700 East Fort Lowell
  Road, Suite 106, Tucson, AZ 85719}

\begin{abstract}

Driven by the incomplete understanding of the formation of gas giant extrasolar planets and of their mass--radius relationship, several ground--based, wide--field photometric campaigns are searching the skies for new transiting extrasolar gas giants. As part of the Trans-atlantic Exoplanet Survey (TrES), in 2003/4 we monitored approximately 30,000 stars ($9 .5\leq V \leq 15.5$) in a $5.7\arcdeg\times5.7\arcdeg$ field in Andromeda with three telescopes over five months. We identified six candidate transiting planets from the stellar light curves. From subsequent follow-up observations, we rejected each of these as an astrophysical false positive, i.e. a stellar system containing an eclipsing binary, whose light curve mimics that of a Jupiter--sized planet transiting a sun--like star. We discuss here the procedures followed by the TrES team to reject false positives from our list of candidate transiting hot Jupiters. We present these candidates as early examples of the various types of astrophysical false postives found in the TrES campaign, and discuss what we learned from the analysis.

\end{abstract}


\section{Introduction}
\label{sec:intro}

There are currently 14 extrasolar planets for which we have measurements of both the planetary radius and mass (\citealp[see][ for a review]{Charbonneau_Brown_Burrows:PPV:2006a}; \citealt{McCullough_Stys_Valenti:apj:2006a, ODonovan_Charbonneau_Mandushev:apjl:2006a, Bakos_Noyes_Kovacs:preprint:2006a, Collier-Cameron_Bouchy_Hebrard:preprint:2006a}). These gas giants have been observed to transit their parent stars, and have supplied new opportunities to study Jupiter--sized exoplanets, in particular their formation and structure. Studies of the visible and infrared atmospheric spectra are possible for the nine nearby ($d<300$\,pc) transiting planets \citep{Charbonneau_Brown_Noyes:apj:2002a, Vidal-Madjar_Lecavelier-des-Etangs_Desert:nat:2003a, Deming_Brown_Charbonneau:apj:2005a, Charbonneau_Allen_Megeath:apj:2005a, Deming_Seager_Richardson:nat:2005a}.
The incident flux from the closeby (\mbox{$\lsim0.05\,\mathrm{A.U.}$}) star on each of these ``hot Jupiters" results in an inflated planetary radius. Current theoretical models that include this stellar insolation can account for the radii of only five of these nine nearby planets. \first\ \citep{Charbonneau_Brown_Latham:apjl:2000a,Henry_Marcy_Butler:apj:2000a}, \tresTwo\ \citep{ODonovan_Charbonneau_Mandushev:apjl:2006a}, \object[HAT-P-1b]{HAT--P--1\,b} \citep{Bakos_Noyes_Kovacs:preprint:2006a}, and \object[WASP-1b]{WASP--1\,b} \citep{Collier-Cameron_Bouchy_Hebrard:preprint:2006a} have radii larger than predicted (\citealp[see][]{Laughlin_Wolf_Vanmunster:apj:2005a} and \citealt{Charbonneau_Brown_Burrows:PPV:2006a} for reviews of the current structural models for insolated hot Jupiters).The sparse sampling and limited understanding of the mass--radius parameter space for extrasolar planets continue to motivate the search for new transiting planets. There are several small aperture wide--field surveys targeting these objects, such as BEST \citep{Rauer_Eisloffel_Erikson:pasp:2004a}, the HAT
network \citep{Bakos_Lazar_Papp:pasp:2002a, Bakos_Noyes_Kovacs:pasp:2004a}, KELT \citep{Pepper_Gould_Depoy:AIP:2004a}, SuperWASP
\citep{Street_Pollaco_Fitzsimmons:ASP:2003a}, Vulcan \citep{Borucki_Caldwell_Koch:pasp:2001a}, and XO \citep{McCullough_Stys_Valenti:pasp:2005a}, as well as deeper surveys like the Optical Gravitational Lensing Experiment (OGLE; \citealt{Udalski_Paczynski_Zebrun:acta:2002a}) that is probing the Galactic disk. 

We are conducting a transit campaign, the \anchor{http://www.astro.caltech.edu/~ftod/tres/}{Trans-atlantic Exoplanet Survey}%
\footnote{\url{http://www.astro.caltech.edu/$\sim$ftod/tres/}}%
\ (TrES), using a network of three 10\,cm telescopes with a wide longitudinal coverage:
\anchor{http://www.astro.caltech.edu/~ftod/tres/sleuth.html}{Sleuth} (located at Palomar Observatory, California; \citealt{ODonovan_Charbonneau_Kotredes:AIP:2004a}), PSST (Lowell
Observatory, Arizona; \citealt{Dunham_Mandushev_Taylor:pasp:2004a}),
and \anchor{http://www.hao.ucar.edu/public/research/stare/stare.html}{STARE}%
\footnote{\url{http://www.hao.ucar.edu/public/research/stare/stare.html}}%
\ (on the isle of Tenerife, Spain; \citealt{Alonso_Deeg_Brown:an:2004a}). The telescopes monitor over several months a $5.7\degr \times 5.7\degr$ field of view containing thousands of nearby bright stars ($9.5\leq V \leq 15.5$), and we examine the light curves of stars with $V\leq14.0$ for repeating eclipses with the short--period, small--amplitude signature of a transiting hot Jupiter. We have discovered two transiting planets so far: \tresOne\ \citep{Alonso_Brown_Torres:apjl:2004a} and \tresTwo\ \citep{ODonovan_Charbonneau_Mandushev:apjl:2006a}. 
In order to find these two transiting planets, we have processed tens of candidates with light curves similar to that of a sun--like star transited by a Jupiter--sized planet. For a typical TrES field at a Galactic latitude of $b\sim15\arcdeg$, we find $\sim10$ candidate transiting planets (\citealp[see, e.g.,][]{Dunham_Mandushev_Taylor:pasp:2004a}). We expect few of these to be true transiting planets, and the remainder to be examples of the various types of astrophysical systems whose light curves can be mistaken for that of a transiting planet (\citealp[see, e.g.,][]{Brown:apjl:2003a, Charbonneau_Brown_Dunham:AIP:2004a}). These are: low-mass dwarfs eclipsing high-mass dwarfs, giant+dwarf eclipsing binaries, and grazing incidence main-sequence eclipsing binaries (with eclipse depths similar to the $\sim1$\% transit depth of a hot Jupiter); and blends, where a faint eclipsing binary and a bright star coincide on the sky or are physically associated, mixing their light  (with the observed eclipse depth reduced to that of a transiting planet). We also encounter occasional photometric false positives, where the transit event is caused by instrumental error, rather than a true reduction in flux from the candidate. \cite{Brown:apjl:2003a} estimates the relative frequency of these astrophysical false positives. For a STARE field in Cygnus, he predicts that from every 25,000 stars observed with sufficient photometric precision to detect a transit, one can expect to identify one star with a transiting planetary companion. However, for this field near the Galactic plane ($b\sim3\degr$), the number of impostor systems identified as candidate planets will outnumber the true detections by an order of magnitude. (The yield of eclipsing systems from such transit surveys depends on the eclipse visibility, which is the fraction of such systems with a given orbital period for which a sufficient number of eclipses could be observed for the system to be detected. This visibility varies with weather conditions and the longitudinal coverage of the telescopes used.) Of the false positives, approximately half are predicted to be eclipsing binaries and half to be blends.

A careful examination of the light curve of a transit candidate may reveal evidence as to nature of the transiting companion. \cite{Seager_Mallen-Ornelas:apj:2003a} present an analytic derivation of the system parameters that can be used to rule out obvious stellar systems. If the light curve demonstrates ellipsoidal variability, this indicates the gravitational influence of a stellar companion \citep{Drake:apj:2003a,Sirko_Paczynski:apj:2003a}. These tests have been used to great effect on the numerous candidates (177 to date) from the OGLE deep--field survey \citep{Drake:apj:2003a, Sirko_Paczynski:apj:2003a, Pont_Bouchy_Melo:aa:2005a, Bouchy_Pont_Melo:aa:2005a}, and candidates from wide--field surveys (\citealp[see, e.g.,][]{Hidas_Ashley_Webb:mnras:2005a, Christian_Pollacco_Skillen:mnras:2006a}).
  
The initial scientific pay-off from each new transiting hot Jupiter comes when an accurate planetary mass and radius have been determined, which can then be used to confront models of planetary structure and formation. These determinations require a high--quality light curve together with a spectroscopic orbit for the host star. For each TrES target field, we follow a procedure of careful examination of each candidate, with follow-up photometry and spectroscopy to eliminate the majority of false positive detections and obtain a high--quality light curve, before committing to the final series of observations with 10-m class telescopes to determine the radial velocity orbit of the candidate planet. This procedure is similar to those discussed by \cite{Charbonneau_Brown_Dunham:AIP:2004a} and \cite{Hidas_Ashley_Webb:mnras:2005a}. Here we discuss our follow-up strategy (\S\ref{sec:elim}) and present the step--by--step results of this procedure for a field in Andromeda, one of the first fields observed by all three nodes of the TrES network. We describe the TrES network observations in \S\ref{sec:obs}, and outline the initial identification of six candidates from the stellar light curves in \S\ref{sec:search}. Based on our follow-up observations of these candidates (\S\ref{sec:followup}), we were able to conclude that each was an astrophysical false positive (\S\ref{sec:reject}).

\section{Follow-up Observations of Planetary Candidates}
\label{sec:elim}

The light curves from small wide--angle telescopes are not of sufficient quality to derive an accurate radius ratio for the purpose of both false positive rejection and planetary modeling, so high--quality follow-up photometric observations with a larger telescope are needed. Recent experience suggests that a photometric accuracy better than 1\,mmag with a time resolution better than 1 minute can be achieved with a meter--class telescope at a good site, and such observations can
deliver radius values good to a few percent and transit times good to 0.2\,min \citep{Holman_Winn_Stanek:preprint:2005a,Holman_Winn_Latham:apj:2006a}. For smaller telescopes, scintillation can limit the photometric precision at this cadence \citep{Young:aj:1967a, Dravins_Lindegren_Mezey:pasp:1998a}. 

The wide--angle surveys by necessity have broad images, typically with FWHM values of $20\arcsec$. Thus there is a significant probability of a chance alignment between a relatively bright star and a fainter eclipsing binary that just happens to be nearby on the sky. Photometric observations with high spatial resolution on a larger telescope can be used to sort out such cases by resolving the eclipsing binary (\citealp[see, e.g.,][]{Charbonneau_Brown_Dunham:AIP:2004a}). In some instances these systems can also be detected using the wide-angle discovery data, by showing that there is differential image motion during the transit events, even though the eclipsing binary is unresolved. Photometry can also be successful in identifying a triple. If the color of the eclipsing binary is different enough from that of the third star, high--quality multi-color light curves will reveal the color--dependent eclipse depths indicative of a such as system (\citealp[see, e.g.,][]{Tingley:aa:2004a, ODonovan_Charbonneau_Torres:apj:2006a}).

A practical problem for this follow-up photometry is that the transiting--planet candidates do not emerge from the wide--angle surveys until late in the observing season, when the observability of the candidates do not permit full coverage of a transit event. With only partial coverage of an event it is difficult to remove systematic drifts across the event, reducing the accuracy of the derived transit depth. Furthermore, full coverage of a transit is important for deriving very accurate transit timings. Without accurate ephemerides, the error in the predicted transit times during the next observing season may be too large to facilitate follow-up photometric observations. The typical duty cycle for a transit is a few hours over a period of a few days, i.e. a few percent. Therefore, if the follow-up photometry does not confirm a transit, the interpretation is ambiguous. The ephemeris may have been too inaccurate, or perhaps the original transit event was a photometric false detection. 

One approach to recovering transits and providing an updated ephemeris for high--quality photometric observations with a larger telescope is to monitor candidates with intermediate sized telescopes, such as TopHAT in the case of the HAT survey \citep{Bakos_Noyes_Kovacs:pasp:2004a}, Sherlock \citep{Kotredes_Charbonneau_Looper:2004a} in the case of TrES, or teams of amateur telescopes \citep{McCullough_Stys_Valenti:apj:2006a} in the case of XO.

A second approach to confirming that a candidate is actually a planet is to obtain very precise radial velocities to see whether the host star undergoes a small reflex motion as expected for a planetary companion. This approach has the advantage that the velocity of the host star varies continuously throughout the orbit, so the observations can be made at any time with only modest attention to the phasing compared to the photometric period.  The ephemeris can then be updated using the velocities, to provide reliable transit predictions for the follow-up photometry. A second advantage is that a spectroscopic orbit is needed anyway to derive the mass of any candidate that proves to be a planet.  The big disadvantage of this approach is that a velocity precision on the order of $10\,\mathrm{m\,s^{-1}}$ is needed, which requires access to a large telescope with an appropriate spectrograph.

For the followup of transiting-planet candidates identified by TrES, we have adopted a strategy designed to eliminate the vast majority of astrophysical false positives with an initial spectroscopic reconnaissance using the Harvard--Smithsonian Center for Astrophysics (CfA) Digital Speedometers \citep{Latham:ASP:1992a} on the 1.5-m Wyeth Reflector at the Oak Ridge Observatory in Harvard, Massachusetts and on the 1.5-m Tillinghast Reflector at the F.~L.~Whipple Observatory (FLWO) on Mount Hopkins, Arizona. We aim to observe candidates spectroscopically during the same season as the discovery photometry. These instruments provide radial velocities good to better than $1\,\mathrm{km\,s^{-1}}$ for stars later than spectral type A that are not rotating too rapidly, and thus can detect motion due to stellar companions with just two or three exposures (\citealp[see, e.g.,][]{Latham:ASP:2003a, Charbonneau_Brown_Dunham:AIP:2004a}). Thus even if the followup is not performed until the target field is almost setting, we can still reject some candidates spectroscopically, even when photometric followup is not useful. For periods of a few days the limiting value for the mass detectable with these instruments is about 5 to 10 Jupiter masses.  

The spectra obtained with these instruments also allow us to characterize the host star. We use a library of synthetic spectra to derive values for the effective temperature and surface gravity, (assuming solar metallicity), and also the line broadening.  In our experience, rotational broadening of more than 10 km/s is a strong hint that the companion is a star, with enough tidal torque to synchronize the rotation of the host star with the orbital motion. Although the gravity determination is relatively crude, with an uncertainty of perhaps 0.5 in $\log{g}$, it is still very useful for identifying those host stars that are clearly giants with $\log{g} \leq 3.0$.  We presume that these stars must be the third member of a system that includes a main-sequence eclipsing binary, either a physical triple or a chance alignment, and we make no further follow-up observations. Our spectroscopic classification of the host star is a first step towards estimating the stellar mass and radius. These estimates, in turn, may be combined with the observed radial velocity variation and light curve to yield estimates of the mass and radius of the companion. 

Although the use of follow-up spectroscopy has the scheduling advantages outlined above, the combination of both spectroscopy and photometry may be needed in the case of a blend. Such a candidate might pass our spectroscopic test as a solitary star with constant radial velocity, if the eclipsing binary of the triple is faint enough relative to the primary star (as was the case for \mbox{\object[GSC 03885-00829]{GSC\,03885--00829}}; \citealt{ODonovan_Charbonneau_Torres:apj:2006a}).

High--precision, high--signal--to--noise spectroscopic observations of the few remaining candidate transiting planets should reveal the mass (and hence true nature) of the transiting companion. However, even after a spectroscopic orbit implying a planetary companion has been derived, care must be taken to show that the velocity shifts are not due to blending with the lines of an eclipsing binary in a triple system (\citealp[e.g.,][]{Mandushev_Torres_Latham:apj:2005a}). It may be hard to see the lines of the eclipsing binary, partly because the eclipsing binary can be quite a bit fainter than the bright third star, and partly because its lines are likely to be much broader due to synchronized rotation.  In some cases it may be possible to extract the velocity of one or both the stars in the eclipsing binary using a technique such as TODCOR \citep{Mandushev_Torres_Latham:apj:2005a}. Combining modeling of the photometric light curve and information from the spectroscopic pseudo orbit for the system can help guide the search for the eclipsing binary lines.  Even if the lines of the eclipsing binary can not be resolved, a bisector analysis of the lines of the third star may reveal subtle shifts that indicate a binary companion.

Follow-up observations with 1-m class telescopes both remove astrophysical false positives from consideration and prepare for the eventual modeling of newly discovered transiting planets. In the case of our field in Andromeda, our followup ruled out all of our planet candidates, and provided us with a variety of false positives to study.

\section{Initial Observations with the TrES Network}
\label{sec:obs}

In August 2003, we selected a new field centered on the guide star \mbox{\object[HD 6811]{HD\,6811}} ($\alpha = 01^{\rm h} 09^{\rm m} 30\fs 13$, $\delta = +47\arcdeg 14\arcmin 30\farcs5$ J2000). We designated this target field as And0, the first TrES field in Andromeda. We observed this field with each of the TrES telescopes. Although the TrES network usually observes concurrently, in this case weather disrupted our observations. Sleuth monitored the field through an SDSS $r$ filter for 42 clear nights between UT 2003 August 27 and October 24. STARE began its observations with a Johnson $R$ filter on UT 2003 September 17, and observed And0 until UT 2004 Jan 13 during 23 photometric nights. PSST went to this field on UT 2003 November 14 and collected Johnson $R$ observations until 2004 January 11, obtaining 19 clear nights. We estimate our recovery rate for transit events should be 100\% for orbital periods $P<6$ days, declining to 70\% for $P=10$ days (see Fig.~\ref{fig:vis}), where here our recovery criterion is the observation of at least half the transit from three distinct transit events. We note that this recovery rate is a necessary but not sufficient criterion to detect transiting planets, since it neglects the signal-to-noise ratio and the detrimental effect of non-Gaussian noise upon it (\citealp[see, e.g.,][]{Gaudi_Seager_Mallen-Ornelas:apj:2005a, Gaudi:apjl:2005a, Pont_Zucker_Queloz:mnras:2006a, Smith_Collier-Cameron_Christian:mnras:2006a, Gaudi_Winn:apj:2007a}). We used an integration time of 90\,s for our exposures. During dark time, we took multicolor photometry (SDSS $g$, $i$, and $z$ for Sleuth and Johnson $B$ and $V$ for PSST and STARE) for stellar color estimates. 

\section{Searching for Transit Candidates}
\label{sec:search}

We have previously described in detail our analysis of TrES data sets in \cite{Dunham_Mandushev_Taylor:pasp:2004a} and \cite{ODonovan_Charbonneau_Torres:apj:2006a}. We summarize here the analysis for this field. We used standard \texttt{IRAF}%
\footnote{IRAF is distributed by the National Optical Astronomy
  Observatories, which are operated by the Association of Universities
  for Research in Astronomy, Inc., under cooperative agreement with
  the National Science Foundation.}%
\ \citep{Tody:1993a} tasks or customized IDL routines to calibrate the images from the three telescopes. For each telescope, we derived a standard list of stars visible in the images from this telescope and computed the corresponding equatorial  coordinates using the Tycho--2 Catalog \citep{Hog_Fabricius_Makarov:aa:2000a}. We applied differential image analysis (DIA) to each of the separate photometric data sets from the three telescopes, using the following pipeline based in part upon \cite{Alard:aas:2000a}. For each star in our standard star lists, we obtained a time series of differential magnitudes with reference to a master image. We produced this master image by combining 20, 18, and 15 of the best--quality interpolated images in our Sleuth, PSST and STARE data sets, respectively. Since small--aperture, wide--field surveys such as TrES often suffer from systematics (caused, for example, by variable atmospheric extinction), we decorrelated the time series of our field stars.

Initially, we examined our Sleuth observations separately, as these dominate the TrES data set, providing 50\% of the data. We binned the Sleuth time series using 9 minute bins to reduce computation time. The rms scatter of the binned data was below 0.015 mag for approximately 7,800 stars.  We searched the time series for periodic transit--like dips using the box--fitting least squares transit--search algorithm (BLS; \citealt*{Kovacs_Zucker_Mazeh:aa:2002a}), which assigns a Signal Detection Efficiency (SDE) statistic to each star, based on the strength of the transit detection. We restricted our search to periods ranging from 0.1 to 10 days. Having sorted the stars in order of decreasing magnitude and decreasing SDE, we visually examined each stellar light curve (phased to the best fit period derived by the algorithm) in turn, until we determined that we could no longer distinguish a transit signal from the noise. We identified six transit candidates (see Tables~\ref{tab:names} and~\ref{tab:transits}, and Fig.~\ref{fig:tresCand}).

\begin{figure}[p]
\begin{center}
\includegraphics[angle=90,scale=0.50]{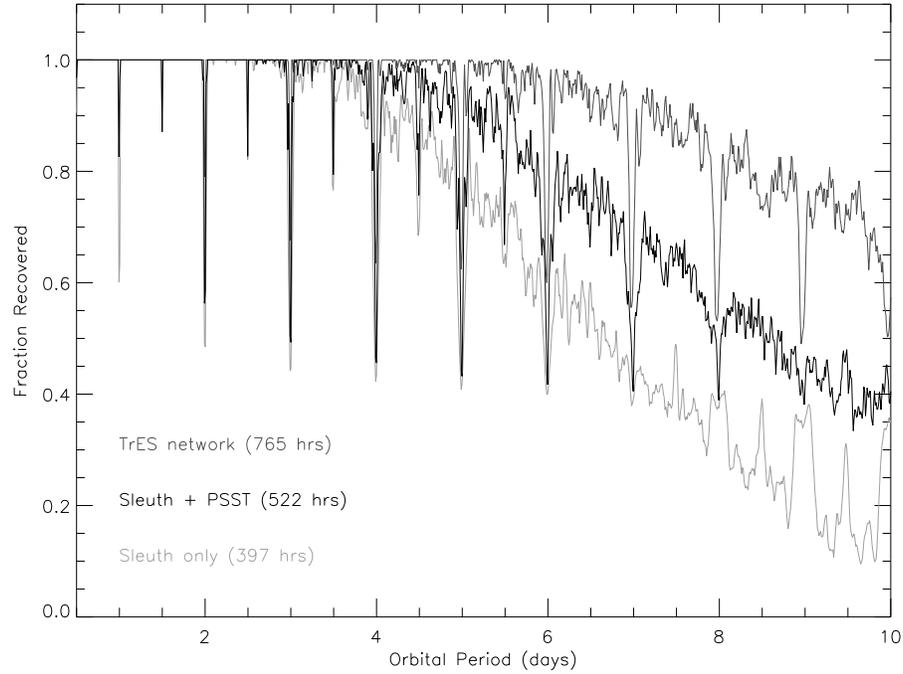}
\end{center}
\caption{Transit visibility plot for the Andromeda field calculated from observations made using Sleuth alone (\textit{light gray}); Sleuth and the PSST (\textit{black}); and all three TrES telescopes (\textit{dark gray}). The fraction of transit signals with a given period identifiable from the data is plotted, assuming a requirement of observing 3 distinct transit events, with coverage of at least half of each individual event. About 80\% of transit events with periods less than 8 days should be recoverable from the TrES observations, whereas the Sleuth observations alone provide 80\% coverage only up to 5 day periods.}
\label{fig:vis}
\end{figure}

\begin{figure}[p]
\begin{center}
\epsscale{1.0}
\plotone{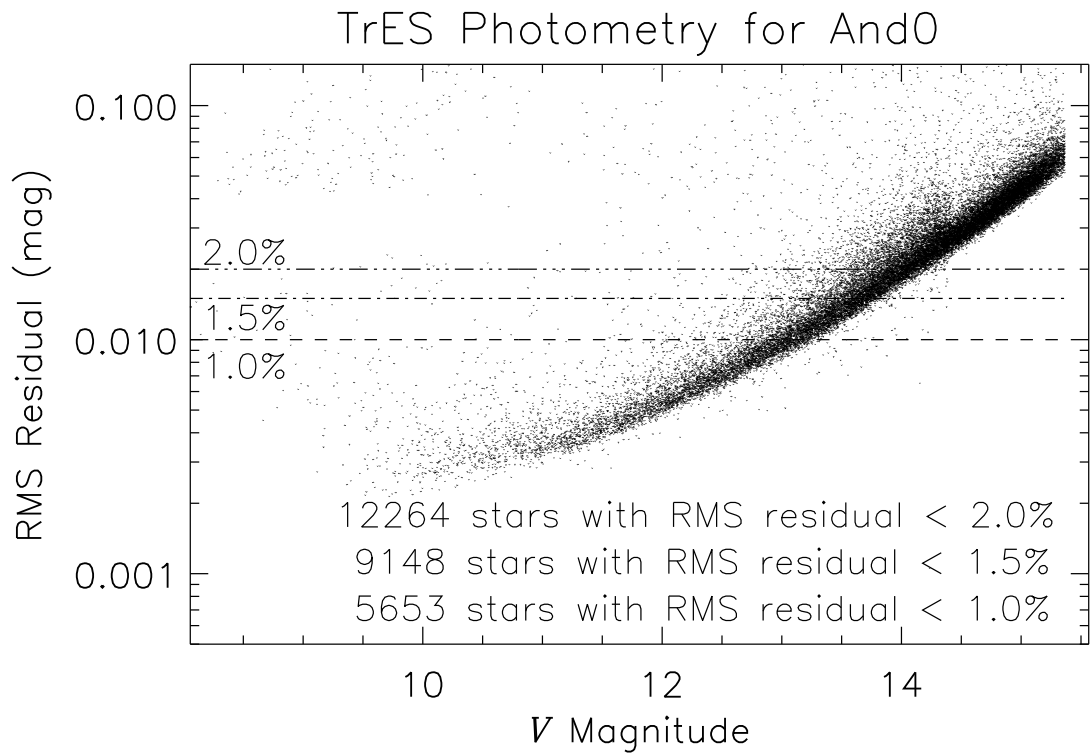}
\end{center}
\caption{The calculated RMS residual of the binned data versus approximate $V$ magnitude for the stars in our TrES And0 data set. The number of stars with RMS below 1\%, 1.5\%, and 2\% are shown.}
\label{fig:rmsplot}
\end{figure}

We then combined the three TrES data sets, which optimized our visibility function (see Fig.~\ref{fig:vis}), and allowed us to confirm the detection of real eclipse events using simultaneous observations from multiple telescopes. We produced the combined TrES data set as follows. For each star in the Sleuth standard star list, we attempted to identify the corresponding stars in the other two lists. We computed the distances between a given Sleuth standard star and the PSST standard stars, and matched the Sleuth star with a PSST star if their angular separation was less than $5\arcsec$ (0.5 pixels). Because of the slight differences in the chosen filter and field of view, some Sleuth stars did not have corresponding PSST stars. We created a new standard star list from the Sleuth and PSST star lists, with only one entry for each pair of matched stars and an entry for each unmatched star. We then repeated this procedure with this new star list and the STARE list to produce the TrES field standard star list. For each star, we then combined the relevant time series, chronologically reordered the data, and binned the data. The rms scatter of the averaged TrES data was below 0.015\,mag for 9,148 stars (see Fig.~\ref{fig:rmsplot}) out of the 29,259 stars in the field. We repeated the BLS transit--search, but did not identify any new candidates. From Figure~\ref{fig:vis}, we can see that a visibility of 80\% had already been achieved for $P<5$\,d for the Sleuth data alone, and the addition of the STARE and PSST data did not significantly increase the detection space for those short periods. The lack of additional candidates with these orbital periods is not surprising. However, for longer periods, the visibility for the TrES network is much better than for the single Sleuth telescope, yet we did not find new candidates with these periods. We proceeded to our follow-up observations of these six candidates with larger telescopes. 

\section{Followup of And0 Candidates}
\label{sec:followup}

Many of the bright stars within our field were also observed as part of other surveys. We identified our candidates in online catalogs and compared these observations with our expectations based on the planet hypothesis. We found Tycho--2 \citep{Hog_Fabricius_Makarov:aa:2000b,Hog_Fabricius_Makarov:aa:2000a} visible ($B_{T}-V_{T}$) colors for two of our candidates, and Two Micron All Sky Survey (2MASS; \citealt{Cutri_Skrutskie_van-Dyk:2003a}) infrared ($J-K$) colors for all six (see Tab.~\ref{tab:catalogs}). We searched the USNO CCD Astrograph Catalog (UCAC2; \citealt{Zacharias_Urban_Zacharias:aj:2004a}) for the proper motions of the stars. All of the candidates display a measurable proper motion, consistent with nearby dwarfs. However, these proper motions were not sufficiently large to rule out distant, high--velocity giants. Finally, we retrieved Digitized Sky Survey%
\footnote{The \anchor{http://archive.stsci.edu/dss/}{Digitized Sky Survey} (\url{http://archive.stsci.edu/dss/}) was produced at the Space Telescope Science Institute under U.S. Government grant NAG W-2166. The images of these surveys are based on photographic data obtained using the Oschin Schmidt Telescope on Palomar Mountain and the UK Schmidt Telescope. The plates were processed into the present compressed digital form with the permission of these institutions.}%
\ (DSS) images of the sky surrounding each candidate to check for possible nearby stars of similar brightness within our PSF radius. None were found. 

We observed the six And0 candidates starting on UT 2004 September 28 using
the Harvard--Smithsonian Center for Astrophysics (CfA) Digital Speedometers \citep{Latham:ASP:1992a}. These spectrographs cover 45\AA\ centered on 5187\AA\, and have a resolution of $8.5\,\mathrm{km\,s^{-1}}$ (a resolving power of $\lambda / \Delta \lambda \approx 35,\!000$). We cross-correlated our spectra against a grid of templates from our library of synthetic spectra to estimate various stellar parameters of our targets and their radial velocities. J. Morse computed this spectral library, using the Kurucz model atmospheres (J.~Morse \& R.~L.~Kurucz, 2004, private communication).
Assuming a solar metallicity, we estimated the effective temperature ($T_{\mathrm{eff}}$), surface gravity ($g$) and rotational velocity ($v \sin{i}$) for each candidate (see Tab.~\ref{tab:catalogs}) from the template parameters that gave the highest average peak correlation value over all the observed spectra. 

For the three candidates with low stellar rotation, $v\sin{i} < 50\,\mathrm{km\,s^{-1}}$, we obtained several spectra over different observing seasons to determine the radial velocity variation of each star. Table~\ref{tab:rv} details our spectroscopic observations. For these slowly rotating candidates, the typical precision for our spectroscopic parameters is $\Delta T_{\mathrm{eff}} = 150\,\mathrm{K}$, $\Delta \log{g} = 0.5$, $\Delta v \sin{i} = 2\,\mathrm{km\,s^{-1}}$ and $\Delta V = 0.5\,\mathrm{km\,s^{-1}}$. The precision of the estimates degrades for stars with large $v\sin{i}$ values or few spectroscopic observations.

We obtained high precision photometry of \tFive\ on UT 2004
December 19 using the Minicam CCD imager at the FLWO 1.2-m telescope on
Mt. Hopkins, Arizona.  Minicam consists of two 2248$\times$4640 pixel
thinned, backside--illuminated Marconi CCDs mounted side--by--side to
span a field of approximately 20.4$\times$23.1 arcminutes bisected by
a narrow gap.  We employed 2$\times$2 pixel binning for an effective
plate scale of $0.6\arcsec\,\mathrm{pixel}^{-1}$ and read out each half CCD
through a separate amplifier.  We offset the telescope to place
\tFive\ centrally on one amplifier region and autoguided on the
field.  We obtained concurrent light curves in 3 filters by cycling
continuously through the SDSS $g$, $r$, and $z$ filters with exposure times
of 90, 45, and 90 seconds respectively.  The seeing was poor
(FWHM $\sim3$--$7\arcsec$) and varied throughout our observations.
Unfortunately, high winds forced us to close the dome during the night
and we obtained only partial coverage of the predicted event.  Late in
the scheduled observations, we re-opened for a short time.  

We subtracted the overscan bias level from each image and divided each
by a normalized flat field constructed from the filtered mean of
twilight sky exposures.  To construct a light curve of \tFive\ and neighboring 
bright stars, we located the stars in each image.  We measured stellar fluxes in a
circular aperture and subtracted the sky as estimated by the median
flux in an annulus centered on the star (iteratively rejecting deviant
sky pixel values).  We used a relatively large 12$\arcsec$ radius
aperture and sky annulus with inner and outer radii of 15$\arcsec$ and 27$\arcsec$
respectively in an effort to reduce systematic errors
arising due to the poor and variable seeing conditions.  We first corrected 
the flux of each star by an amount proportional to its
airmass in each exposure by using extinction coefficients for each
filter based on previous experience with Minicam photometry.  Second,
we selected a group of bright, uncrowded stars near \tFive\ as
potential comparison stars.  In each exposure, we calculated the mean flux
of the comparison stars weighted according to brightness.  We assumed
that any variations in this mean flux represented extinction in each image
and used them to apply corrections to each light curve.  We then
inspected by eye the light curve of each comparison star, and fit the light curves to
models of constant brightness to find chi-squared statistics. We
removed from our group of comparison stars any star that showed significant
variations. We recalculated the extinction corrections iteratively in
this manner until we achieved no variation in the comparison star light curves. 
We accepted 29, 32, and 4 comparison stars for the $g$, $r$, and $z$ band light curves respectively. Finally, we normalized the flux in the light curves of \tFive\ with respect to the out--of--eclipse data.

\section{Rejecting False Positive Detections}
\label{sec:reject}

Based on our detailed investigations of the candidates, we eliminated each And0 candidate as follows. 

\begin{figure}[p]
\begin{center}
\epsscale{1.0}
\plotone{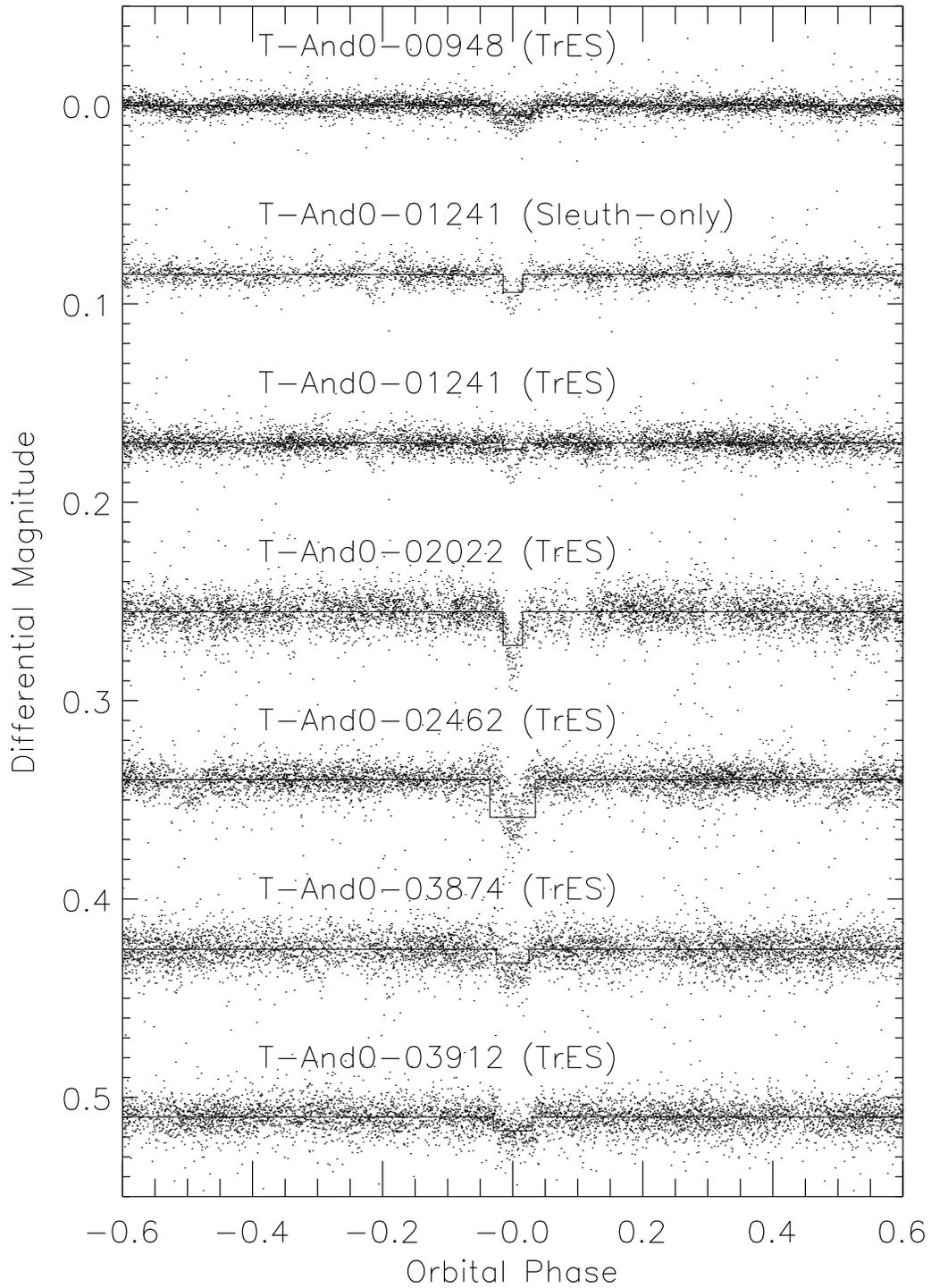}
\end{center}
\caption{The light curves of the six TrES candidates from the And0 field in Andromeda. The labels denote the source of the light curve. The timeseries have been phased to the best--fit period identified by the box--fitting algorithm of \cite{Kovacs_Zucker_Mazeh:aa:2002a} using the TrES data, with the exception of \mbox{T--And0--01241}, whose period whose derived from the Sleuth data. The transit event is not present in the data from the other telescopes gathered at the same orbital phase.}
\label{fig:tresCand}
\end{figure}

In the case of \tOne, the TrES light curve (Fig.~\ref{fig:tresCand}) shows a secondary eclipse. The low surface gravity ($\log{g}=3.0$) is that of a distant giant star, consistent with the red color $J-K=0.60$ (since the majority of stars with $J-K>0.5$ are expected to be giants \citealp[see, e.g.,][]{Brown:apjl:2003a}). There was no observed variation in the radial velocity of this candidate ($v_r=-28.82\,\mathrm{km\,s^{-1}}$). \tOne\ is most likely the primary star of a diluted triple system. 

\begin{figure}[p]
     \centering
          \includegraphics[width=.45\textwidth]{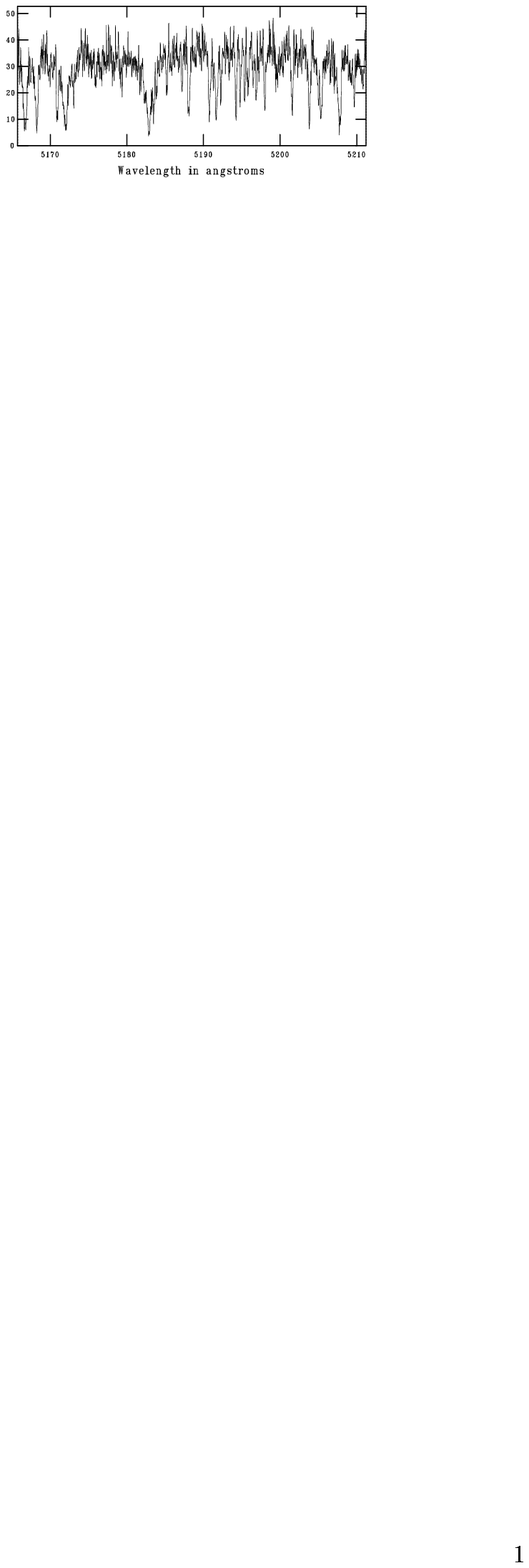}
	\hspace{1cm}	
          \includegraphics[width=.45\textwidth]{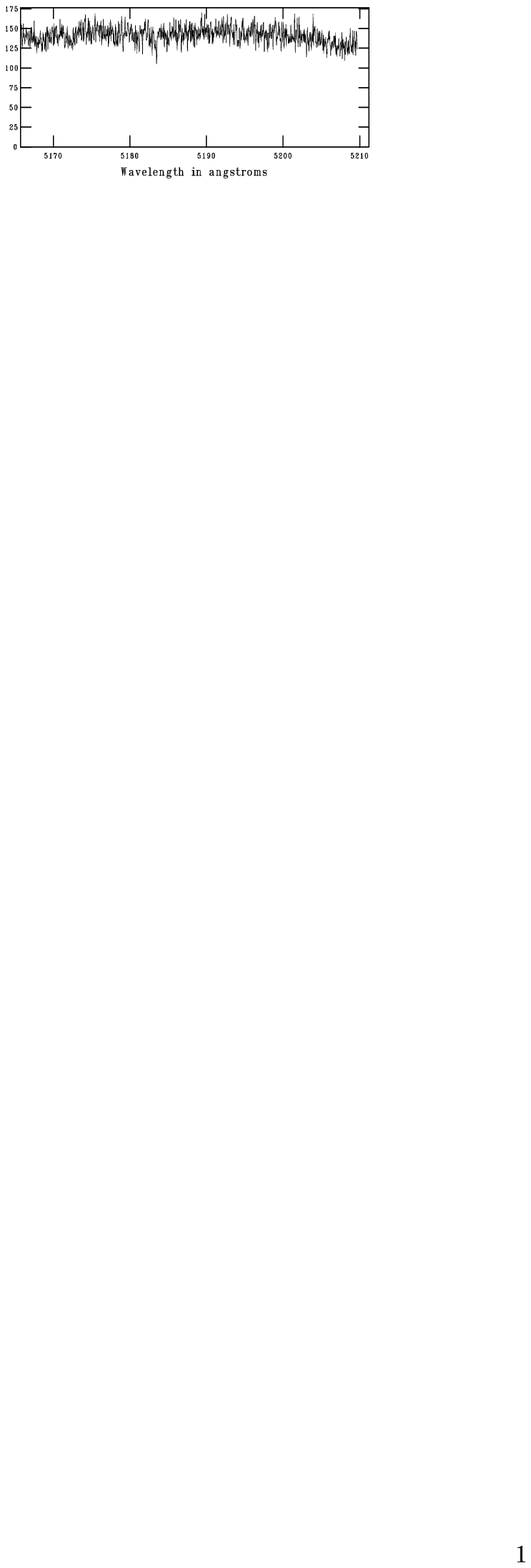}\\
          \includegraphics[width=.45\textwidth]{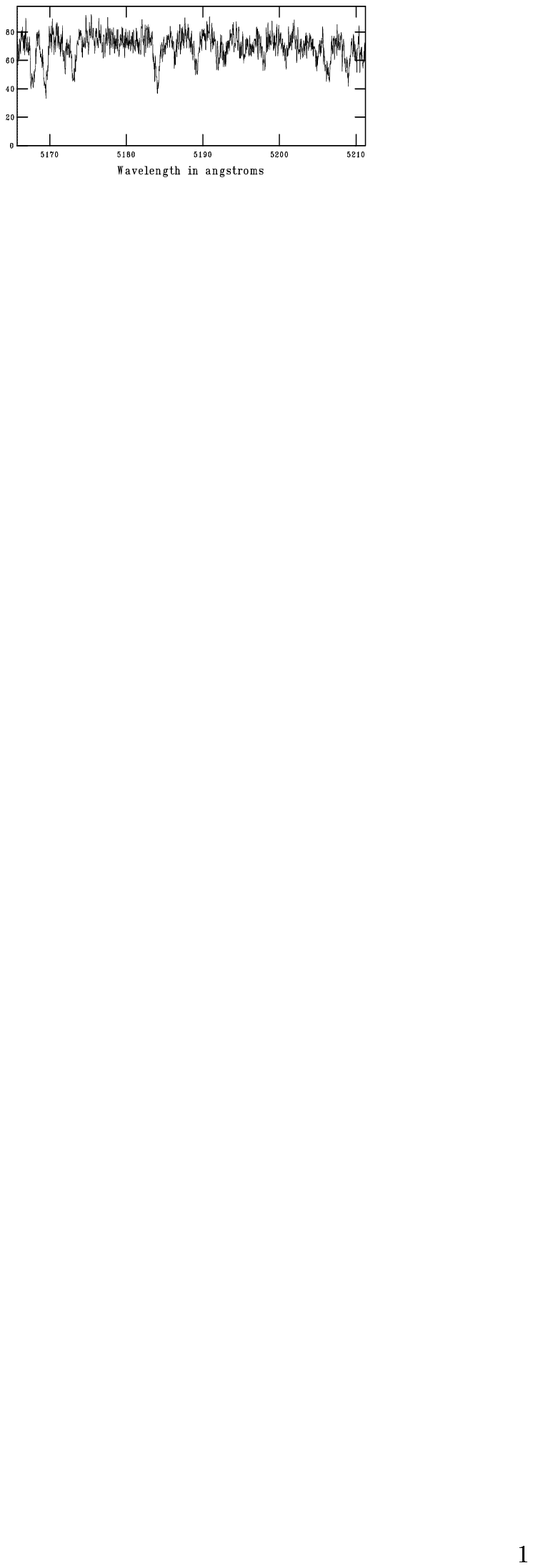}
          \hspace{1cm}
          \includegraphics[width=.45\textwidth]{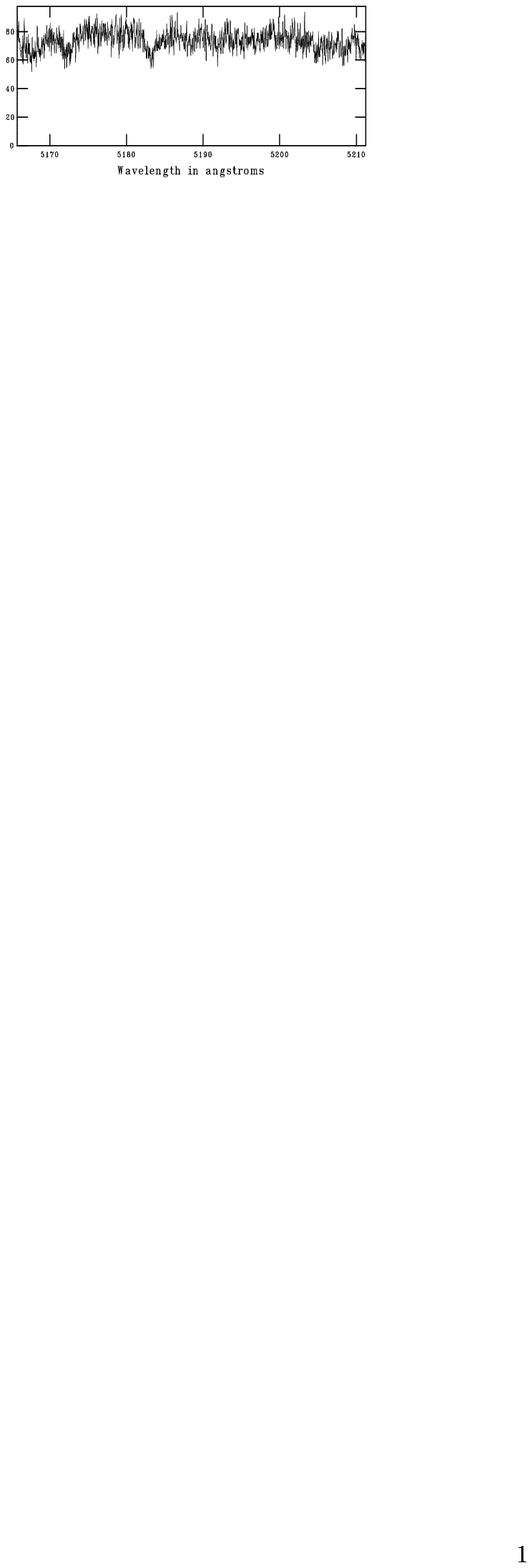}\\
          \includegraphics[width=.45\textwidth]{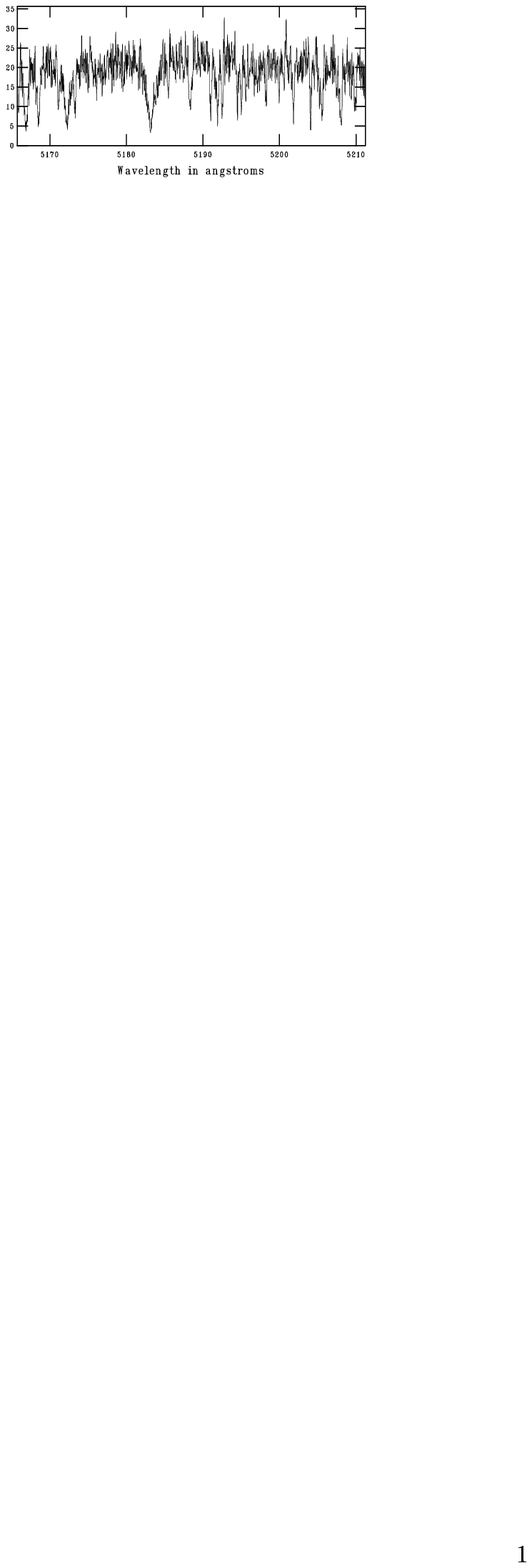}
          \hspace{1cm}
          \includegraphics[width=.45\textwidth]{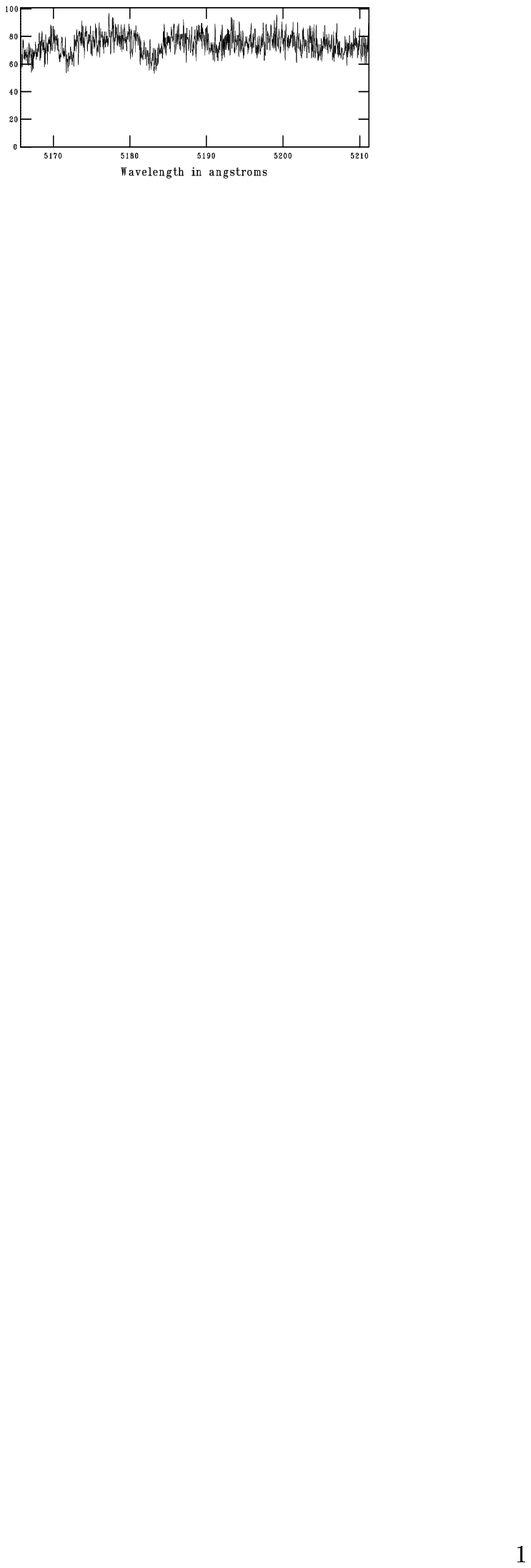}
     \caption{
Sample spectra of the And0 TrES candidates obtained with the CfA Digital Speedometers on the FLWO 1.5-m telescope: \mbox{T--And0--00948} (\textit{upper--left}) has the low surface gravity of a giant star; \mbox{T--And0--01241} (\textit{upper--right}) has the featureless spectrum of an A--type star; \mbox{T--And0--02022} (\textit{center--left}) is an evolved F dwarf; \mbox{T--And0--02462} (\textit{center--right}) and \mbox{T--And0--03912} (\textit{lower--right}) display broadened lines due to the rapid rotation of these stars; \mbox{T--And0--03874} (\textit{lower--left}) is an early K--type dwarf.}
\label{fig:spectra}
\end{figure}

Upon further examination of the individual light curves for \tTwo, we noticed that only Sleuth had observed transit events for this system. Neither PSST nor STARE had observed this field during the time of the Sleuth transit events, preventing a comparison of the light curves. Based on the Sleuth data, we predicted the times of transits during the entire TrES And0 campaign. STARE did observe \tTwo\ at a time at which it was predicted to transit but did not observe the transit. It was therefore possible that this was a photometric false positive. However, we did not pursue this further, as we obtained sufficient evidence from the follow-up spectroscopy to discount the system. A dwarf with $\log{g}=4.5$, the star has the high effective temperature ($T_{\mathrm{eff}}=9500$\,K) and blue colors ($J-K=0.01$) of an early A star. Figure~\ref{fig:spectra}$b$ shows the nearly featureless spectrum of this star. For such a large star with a radius $R\sim2.7\,R_{\sun}$, the observed transit depth of 0.9\% indicates a non--planetary size ($R=2.5\,R_{\mathrm{Jup}}$) for the eclipsing body. 

\begin{figure}[p]
\epsscale{0.65}
\begin{center}
\plotone{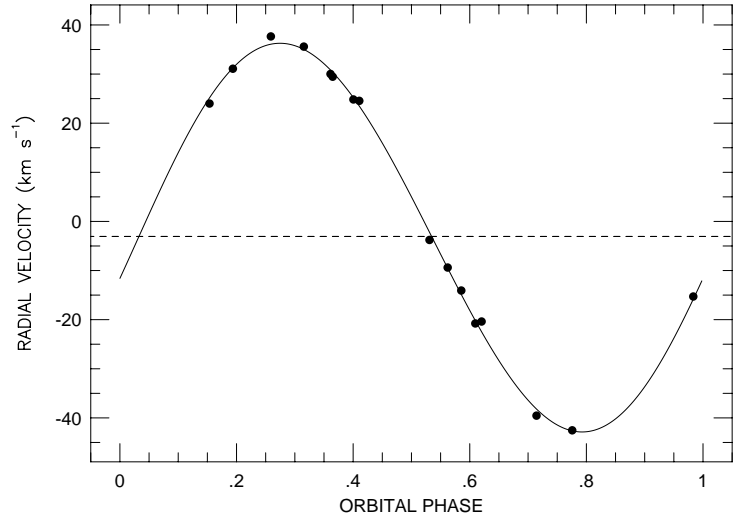}
\end{center}
\caption{
The radial velocity orbit of \mbox{T--And0--02022} as determined from our Digital Speedometer spectra. This system was quickly rejected as a candidate transiting planet after the large radial velocity variation was determined from initial spectroscopic observations. Additional epochs produced a precise orbit, with an eccentricity of $e\sim0.03$. An initial mass estimate of $0.5\,\mathrm{M}_{\sun}$ was derived for the companion. Future photometry will lead to more precise mass determination for both component stars (see text for a discussion).}
\label{fig:rvorbit}
\end{figure}

The radial velocity of \tThree\ varies with an amplitude corresponding to a stellar--mass companion. We determined this system to be an eclipsing binary, comprised of a slightly evolved F dwarf and an M dwarf. This system has a mass function $f(M)=0.0304\pm0.0013\,\mathrm{M}_{\sun}$ and an eccentricity of $0.027\pm0.014$. Assuming an orbital inclination of $i\sim90\degr$, and a mass of $1.6\,\mathrm{M}_{\sun}$ consistent with the effective temperature ($T_{\mathrm{eff}}=7000$\,K), we estimated the mass of the companion to be $m=0.5\,\mathrm{M}_{\sun}$. Figure~\ref{fig:rvorbit} shows the radial velocity orbit. The circular orbit of \tThree\ allows us to constrain the stellar radius, independent of our derived spectral type and luminosity class. The circular orbit implies orbital synchronization and orbital--rotation axes alignment, since circularization has the longest timescale of these processes. (This should apply except for very low mass secondaries; \citealp[see][]{Hut:aa:1981a} for a derivation of these timescales for close binary systems and a comparison for different binary mass ratios and moments of inertia of the primary star.) We can therefore assume that the stellar rotation period is the same as the orbital period of $4.7399$d. We then use the observed rotational broadening ($v\sin{i}=22\,\mathrm{km\,s^{-1}}$) to estimate the radius of the star to be $2.0\,R_{\sun}$. Future photometric observations of this systems with KeplerCam \citep{Szentgyorgyi_Geary_Latham:AAS:2005a} are planned to more precisely measure the eclipse depth, and to derive the radius and true mass of each star in this binary (Fernandez~et~al., in preparation).

\tFour\ is a rapid rotator with $v\sin{i}=77\,\mathrm{km\,s^{-1}}$, and displays rotationally broadened lines (see Figure~\ref{fig:spectra}$d$), limiting the possibility of detecting the radial velocity variation caused by a planet. Regardless, we implied a binary nature for this candidate using the combined TrES data, which proved essential in identifying this false positive. Our initial Sleuth light curve of \tFour\ showed no evidence of a stellar--mass companion, and we derived a BLS best--fit period of 1.5347\,d. Upon examining the TrES light curve of \tFour\ (see Fig.~\ref{fig:tresCand}), we noticed a secondary eclipse. Also, the BLS best--fit period for our TrES observations of \tFour\ is twice that derived from the Sleuth data alone. When we re-examined the Sleuth data, phased to the TrES period, the secondary eclipse is not visible, due to inadequate coverage at that phase. This resulted in a derived period for the Sleuth data half that of the true period. 

\begin{figure}[p]
\epsscale{1.0}
\begin{center}
\includegraphics[angle=90,scale=0.50]{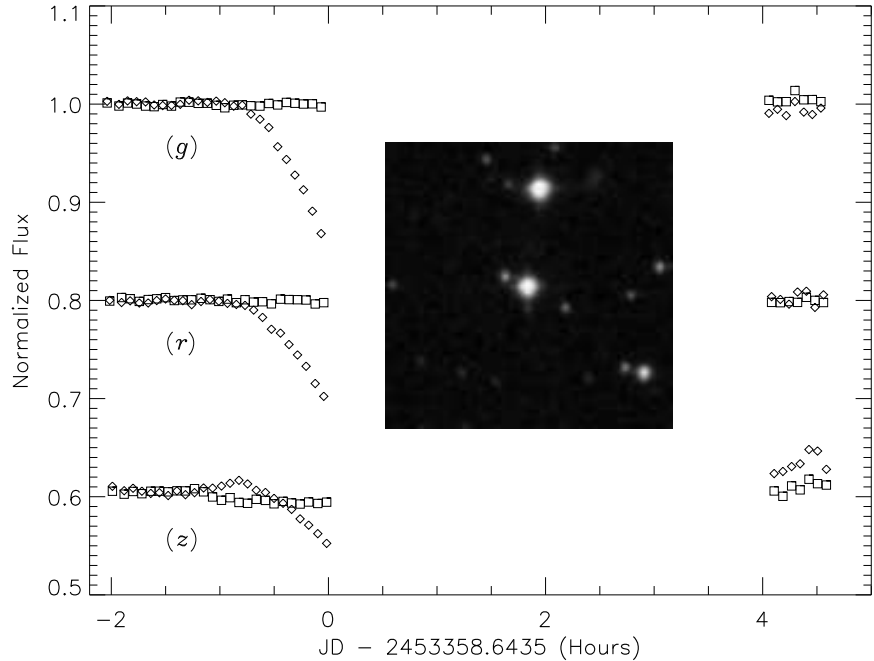}
\end{center}
\caption{
Follow-up $g$, $r$ and $z$ band photometry with the FLWO 1.2-m telescope of \mbox{T--And0--03874} (\textit{squares}) and a neighboring star (\textit{diamonds}), designated \mbox{T--And0--02943}, that lies $45\arcsec$ away. (The inset $2\arcmin \times 2\arcmin$ Digitized Sky Survey image shows \mbox{T--And0--03874} at the center and \mbox{T--And0--02943} to the north.) The flux from each star has been normalized using the out--of--eclipse data, and an offset  applied for the purpose of plotting. Inclement weather prevented complete coverage of the predicted eclipse event. Nevertheless, it is evident that \mbox{T--And0--02943} displays a deep ($>14$\%) eclipse, whereas the flux from \mbox{T--And0--03874} is constant. The observed apparent transits of \mbox{T--And0--03874} were the result of the blending of the light from these two systems. With a FWHM for \mbox{T--And0--02943} of $\sim2.5$\,pixels ($\sim25\arcsec$), some of the light from this star was within the photometric aperture radius ($3$\,pixels; $\sim30\arcsec$) of \mbox{T--And0--03874}.}
\label{fig:blend}
\end{figure}

\begin{figure}[p]
\epsscale{1.0}
\begin{center}
\includegraphics[angle=90,scale=0.50]{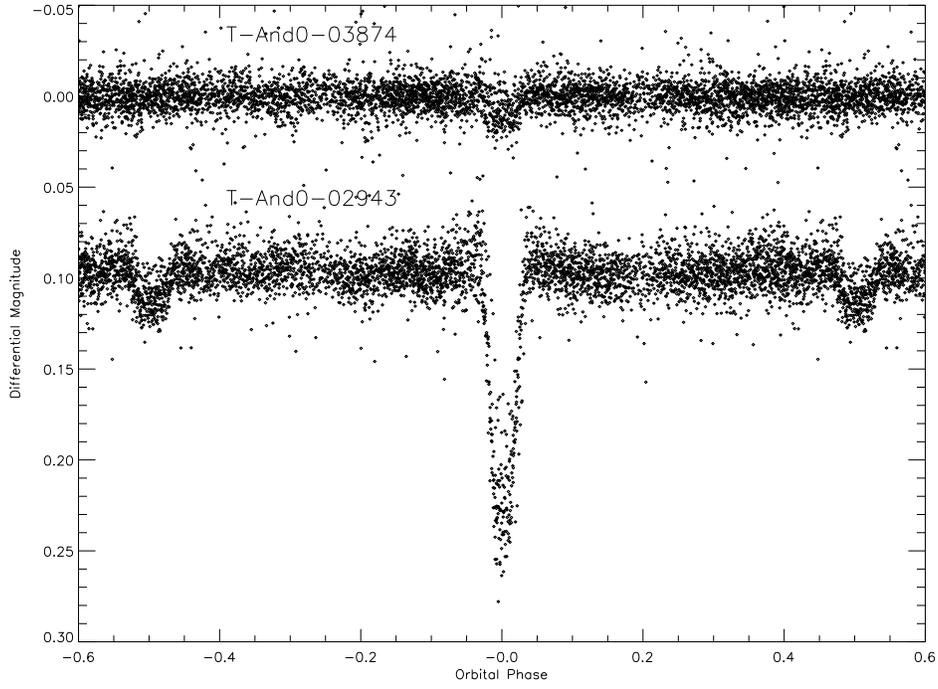}
\end{center}
\caption{TrES photometry of \mbox{T--And0--03874} and the neighboring eclipsing binary \mbox{T--And0--02943}, phased to the best--fit period (2.6540d) for \mbox{T--And0--03874} derived by the box--fitting algorithm. Both stars show eclipses with the same orbital period and epoch. The $\sim1$\% transit--like event detected in the light curve of \mbox{T--And0--03874} was the result of the blending of light from this star and from \mbox{T--And0--02943}, which lies $45\arcsec$ away, comparable to the $30\arcsec$ radius of the TrES PSF. }
\label{fig:compblend}
\end{figure}

The red color of \tFive\ ($J-K=0.66$) and the effective temperature ($T_{\mathrm{eff}}=5500$\,K) calculated from the spectrum shown in Figure~\ref{fig:spectra}$e$ are consistent with an early K--type star. The radial velocity of \tFive\ was observed to remain constant at $-15.53\,\mathrm{km\,s^{-1}}$. However, the low estimated surface gravity ($\log{g}=3.5$) suggested this star is a giant star and part of a diluted triple. The photometric follow-up (see Fig.~\ref{fig:blend}) failed to recover transits of \tFive, but did observe a nearby eclipsing binary \mbox{T--And0--02943} undergoing a deep eclipse at the predicted transit time (also shown in Fig.~\ref{fig:blend}). When we examined our TrES observations for \mbox{T--And0--02943}, we saw that the period of this eclipsing binary was that originally derived for \tFive, namely 2.654 days. This eclipsing system lies $45\arcsec$ away from \tFive, comparable to the PSF radius of our TrES aperture photometry ($30\arcsec$; 3 pixels). The angular resolution ($\sim1\arcsec\,\mathrm{pixel}^{-1}$) of the 1.2-m photometry is higher than that of the original TrES photometry ($9.9\arcsec\,\mathrm{pixel}^{-1}$); hence the light from these two systems is blended in our TrES observations (see Fig.~\ref{fig:compblend}) but is resolved in the 1.2-m photometry. 

Finally, the observed parameters for \tSix\ ($\log{g}=3.5$, $T_{\mathrm{eff}}=7750$\,K, $J-K=0.25$) again imply a large stellar radius ($R\sim1.8\,R_{\sun}$). In order to produce the 0.9\% transit, a companion radius of $1.5\,R_{\mathrm{Jup}}$ is required. This is consistent with the large radii of the transiting planets
 \object[HAT-P-1b]{HAT--P--1\,b} \citep{Bakos_Noyes_Kovacs:preprint:2006a}, and \object[WASP-1b]{WASP--1\,b} \citep{Charbonneau_Winn_Everett:preprint:2007a},
although we have not ruled out the possibility of a blend. Regardless, the star \tSix\ is rapidly rotating ($v\sin{i}=88\,\mathrm{km\,s^{-1}}$; see Fig.~\ref{fig:spectra}$f$), making precise radial-velocity followup extremely difficult.

\section{Discussion}
\label{sec:discuss}

The Trans-atlantic Exoplanet Survey monitors $\sim30,000$ stars each year from which we identify $\sim30$ stars whose light curves show periodic eclipses consistent with the passage of a Jupiter--sized planet in front of a sun--like star. In order to eliminate astrophysical false positives, we have established a procedure of multi--epoch photometric and moderate--precision spectroscopic follow-up. Surviving candidates are optimal targets for high--precision multi--epoch radial velocity measurements that will yield the masses of planetary companions.

The TrES field in Andromeda was one of the first of our fields for which we combined the data from the three TrES telescopes. We have demonstrated here the benefit of this: the improved transit visibility, the increase in number of stars with low RMS residual, and the confirmation of transit events observed by different telescopes. It has also provided us with several examples of the astrophysical false positives that will be encountered in any ground--based, wide--field transit survey, all of which were rejected as a result of follow-up observations. Based on our experience with these false positives, we refined the criteria by which we identify TrES planet candidates. For each new transit candidate, we now search our dataset for a nearby eclipsing binary with a similar period as derived by the BLS algorithm, in order to check for blends. We also examine the transit light curve phased using integer multiples or fractions of the BLS orbital period. 

In comparison to other, more subtle examples (such as those discussed by \citealt{ODonovan_Charbonneau_Torres:apj:2006a}, \citealt{Torres_Konacki_Sasselov:apj:2004b}, and \citealt{Mandushev_Torres_Latham:apj:2005a}), these transit candidates were easily identified as false positives. These examples demonstrate the need for both spectroscopic and photometric follow-up of transit candidates, which may be accomplished with the 1-m class telescopes, on which time is readily available. This ensures that the precious resource of 10-m spectroscopy is used efficiently. 

\acknowledgments	

FTOD and DC thank Lynne Hillenbrand for her supervision of this thesis work.
This material is based upon work supported by the National Aeronautics and Space
Administration under grants NNG05GJ29G, NNG05GI57G, NNH05AB88I, and NNG04LG89G, issued through the Origins of Solar Systems Program. We acknowledge support for this work from the \textit{Kepler} mission via NASA Cooperative Agreement NCC2--1390. This research has made use of the SIMBAD database, operated at CDS, Strasbourg, France, and NASA's Astrophysics Data System Bibliographic Services.  This publication also utilizes data products from the Two Micron All Sky Survey, which is a joint project of the University of Massachusetts and the Infrared Processing and Analysis Center/California Institute of Technology, funded by the National Aeronautics and Space Administration and the National Science Foundation. 

{\it Facilities:} PO:Sleuth, FLWO:1.2m, FLWO:1.5m

\bibliographystyle{apj}
\bibliography{apjmnemonic,mybib.planets}

\begin{thebibliography}{}

\bibitem[\protect\citeauthoryear{{Alard}}{{Alard}}{2000}]{Alard:aas:2000a}
{Alard}, C. 2000, A\&AS, 144, 363

\bibitem[\protect\citeauthoryear{{Alonso} et~al.}{{Alonso}
  et~al.}{2004a}]{Alonso_Brown_Torres:apjl:2004a}
{Alonso}, R., et~al. 2004a, ApJ, 613, L153

\bibitem[\protect\citeauthoryear{{Alonso} et~al.}{{Alonso}
  et~al.}{2004b}]{Alonso_Deeg_Brown:an:2004a}
{Alonso}, R., {Deeg}, H.~J., {Brown}, T.~M.,  \& {Belmonte}, J.~A. 2004b,
  Astron. Nachr., 325, 594

\bibitem[\protect\citeauthoryear{{Bakos} et~al.}{{Bakos}
  et~al.}{2004}]{Bakos_Noyes_Kovacs:pasp:2004a}
{Bakos}, G., {Noyes}, R.~W., {Kov{\'a}cs}, G., {Stanek}, K.~Z., {Sasselov},
  D.~D.,  \& {Domsa}, I. 2004, PASP, 116, 266

\bibitem[\protect\citeauthoryear{{Bakos} et~al.}{{Bakos}
  et~al.}{2002}]{Bakos_Lazar_Papp:pasp:2002a}
{Bakos}, G.~{\'A}., {L{\'a}z{\'a}r}, J., {Papp}, I., {S{\'a}ri}, P.,  \&
  {Green}, E.~M. 2002, PASP, 114, 974

\bibitem[\protect\citeauthoryear{{Bakos} et~al.}{{Bakos}
  et~al.}{2006}]{Bakos_Noyes_Kovacs:preprint:2006a}
{Bakos}, G.~A., et~al. 2006, ApJ, in press (astro-ph/0609369)

\bibitem[\protect\citeauthoryear{{Borucki} et~al.}{{Borucki}
  et~al.}{2001}]{Borucki_Caldwell_Koch:pasp:2001a}
{Borucki}, W.~J., {Caldwell}, D., {Koch}, D.~G., {Webster}, L.~D., {Jenkins},
  J.~M., {Ninkov}, Z.,  \& {Showen}, R. 2001, PASP, 113, 439

\bibitem[\protect\citeauthoryear{{Bouchy} et~al.}{{Bouchy}
  et~al.}{2005}]{Bouchy_Pont_Melo:aa:2005a}
{Bouchy}, F., {Pont}, F., {Melo}, C., {Santos}, N.~C., {Mayor}, M., {Queloz},
  D.,  \& {Udry}, S. 2005, A\&A, 431, 1105

\bibitem[\protect\citeauthoryear{{Brown}}{{Brown}}{2003}]{Brown:apjl:2003a}
{Brown}, T.~M. 2003, ApJ, 593, L125

\bibitem[\protect\citeauthoryear{{Charbonneau} et~al.}{{Charbonneau}
  et~al.}{2005}]{Charbonneau_Allen_Megeath:apj:2005a}
{Charbonneau}, D., et~al. 2005, ApJ, 626, 523

\bibitem[\protect\citeauthoryear{{Charbonneau} et~al.}{{Charbonneau}
  et~al.}{2006}]{Charbonneau_Brown_Burrows:PPV:2006a}
{Charbonneau}, D., {Brown}, T.~M., {Burrows}, A.,  \& {Laughlin}, G. 2006, in
  Protostars and Planets V, ed. B.~{Reipurth}, D.~{Jewitt}, \& K.~{Keil}
  (Tucson: Univ. of Arizona Press), in press, (astro-ph/0603376)

\bibitem[\protect\citeauthoryear{{Charbonneau} et~al.}{{Charbonneau}
  et~al.}{2004}]{Charbonneau_Brown_Dunham:AIP:2004a}
{Charbonneau}, D., {Brown}, T.~M., {Dunham}, E.~W., {Latham}, D.~W., {Looper},
  D.~L.,  \& {Mandushev}, G. 2004, in AIP Conf. Proc. 713: The Search for Other
  Worlds, ed. S.~S. {Holt} \& D.~{Deming} (New York: AIP), 151

\bibitem[\protect\citeauthoryear{{Charbonneau} et~al.}{{Charbonneau}
  et~al.}{2000}]{Charbonneau_Brown_Latham:apjl:2000a}
{Charbonneau}, D., {Brown}, T.~M., {Latham}, D.~W.,  \& {Mayor}, M. 2000, ApJ,
  529, L45

\bibitem[\protect\citeauthoryear{{Charbonneau} et~al.}{{Charbonneau}
  et~al.}{2002}]{Charbonneau_Brown_Noyes:apj:2002a}
{Charbonneau}, D., {Brown}, T.~M., {Noyes}, R.~W.,  \& {Gilliland}, R.~L. 2002,
  ApJ, 568, 377

\bibitem[\protect\citeauthoryear{{Charbonneau} et~al.}{{Charbonneau}
  et~al.}{2007}]{Charbonneau_Winn_Everett:preprint:2007a}
{Charbonneau}, D., {Winn}, J.~N., {Everett}, M.~E., {Latham}, D.~W., {Holman},
  M.~J., {Esquerdo}, G.~A.,  \& {O'Donovan}, F.~T. 2007, ApJ, in press
  (astro-ph/0610589)

\bibitem[\protect\citeauthoryear{{Christian} et~al.}{{Christian}
  et~al.}{2006}]{Christian_Pollacco_Skillen:mnras:2006a}
{Christian}, D.~J., et~al. 2006, MNRAS, 372, 1117

\bibitem[\protect\citeauthoryear{{Collier Cameron} et~al.}{{Collier Cameron}
  et~al.}{2006}]{Collier-Cameron_Bouchy_Hebrard:preprint:2006a}
{Collier Cameron}, A., et~al. 2006, MNRAS, submitted (astro-ph/0609688)

\bibitem[\protect\citeauthoryear{{Cutri} et~al.}{{Cutri}
  et~al.}{2003}]{Cutri_Skrutskie_van-Dyk:2003a}
{Cutri}, R.~M., et~al. 2003, 2MASS All Sky Catalog of Point Sources (Amherst:
  Univ. Massachusetts Press), http://irsa.ipac.caltech.edu/applications/Gator

\bibitem[\protect\citeauthoryear{{Deming} et~al.}{{Deming}
  et~al.}{2005a}]{Deming_Brown_Charbonneau:apj:2005a}
{Deming}, D., {Brown}, T.~M., {Charbonneau}, D., {Harrington}, J.,  \&
  {Richardson}, L.~J. 2005a, ApJ, 622, 1149

\bibitem[\protect\citeauthoryear{{Deming} et~al.}{{Deming}
  et~al.}{2005b}]{Deming_Seager_Richardson:nat:2005a}
{Deming}, D., {Seager}, S., {Richardson}, L.~J.,  \& {Harrington}, J. 2005b,
  Nature, 434, 740

\bibitem[\protect\citeauthoryear{{Drake}}{{Drake}}{2003}]{Drake:apj:2003a}
{Drake}, A.~J. 2003, ApJ, 589, 1020

\bibitem[\protect\citeauthoryear{{Dravins} et~al.}{{Dravins}
  et~al.}{1998}]{Dravins_Lindegren_Mezey:pasp:1998a}
{Dravins}, D., {Lindegren}, L., {Mezey}, E.,  \& {Young}, A.~T. 1998, PASP,
  110, 610

\bibitem[\protect\citeauthoryear{{Dunham} et~al.}{{Dunham}
  et~al.}{2004}]{Dunham_Mandushev_Taylor:pasp:2004a}
{Dunham}, E.~W., {Mandushev}, G.~I., {Taylor}, B.~W.,  \& {Oetiker}, B. 2004,
  PASP, 116, 1072

\bibitem[\protect\citeauthoryear{{Gaudi}}{{Gaudi}}{2005}]{Gaudi:apjl:2005a}
{Gaudi}, B.~S. 2005, ApJ, 628, L73

\bibitem[\protect\citeauthoryear{{Gaudi}, {Seager}, \&
  {Mallen-Ornelas}}{{Gaudi}
  et~al.}{2005}]{Gaudi_Seager_Mallen-Ornelas:apj:2005a}
{Gaudi}, B.~S., {Seager}, S.,  \& {Mallen-Ornelas}, G. 2005, ApJ, 623, 472

\bibitem[\protect\citeauthoryear{{Gaudi} \& {Winn}}{{Gaudi} \&
  {Winn}}{2007}]{Gaudi_Winn:apj:2007a}
{Gaudi}, B.~S.,  \& {Winn}, J.~N. 2007, ApJ, 655, 550

\bibitem[\protect\citeauthoryear{{Henry} et~al.}{{Henry}
  et~al.}{2000}]{Henry_Marcy_Butler:apj:2000a}
{Henry}, G.~W., {Marcy}, G.~W., {Butler}, R.~P.,  \& {Vogt}, S.~S. 2000, ApJ,
  529, L41

\bibitem[\protect\citeauthoryear{{Hidas} et~al.}{{Hidas}
  et~al.}{2005}]{Hidas_Ashley_Webb:mnras:2005a}
{Hidas}, M.~G., et~al. 2005, MNRAS, 360, 703

\bibitem[\protect\citeauthoryear{{H{\o}g} et~al.}{{H{\o}g}
  et~al.}{2000a}]{Hog_Fabricius_Makarov:aa:2000b}
{H{\o}g}, E., et~al. 2000a, A\&A, 357, 367

\bibitem[\protect\citeauthoryear{{H{\o}g} et~al.}{{H{\o}g}
  et~al.}{2000b}]{Hog_Fabricius_Makarov:aa:2000a}
{H{\o}g}, E., et~al. 2000b, A\&A, 355, L27

\bibitem[\protect\citeauthoryear{{Holman} et~al.}{{Holman}
  et~al.}{2006}]{Holman_Winn_Latham:apj:2006a}
{Holman}, M.~J., et~al. 2006, ApJ, 652, 1715

\bibitem[\protect\citeauthoryear{{Holman} et~al.}{{Holman}
  et~al.}{2005}]{Holman_Winn_Stanek:preprint:2005a}
{Holman}, M.~J., {Winn}, J.~N., {Stanek}, K.~Z., {Torres}, G., {Sasselov},
  D.~D., {Allen}, R.~L.,  \& {Fraser}, W. 2005, ApJ, in press
  (astro-ph/0506569)

\bibitem[\protect\citeauthoryear{{Holt} \& {Deming}}{{Holt} \&
  {Deming}}{2004}]{Holt_Deming:AIP:2004a}
{Holt}, S.~S.,  \& {Deming}, D., ed. 2004, {The Search for Other Worlds}

\bibitem[\protect\citeauthoryear{{Hut}}{{Hut}}{1981}]{Hut:aa:1981a}
{Hut}, P. 1981, A\&A, 99, 126

\bibitem[\protect\citeauthoryear{{Kotredes} et~al.}{{Kotredes}
  et~al.}{2004}]{Kotredes_Charbonneau_Looper:2004a}
{Kotredes}, L., {Charbonneau}, D., {Looper}, D.~L.,  \& {O'Donovan}, F.~T.
  2004, in AIP Conf. Proc. 713: The Search for Other Worlds, ed. S.~S. {Holt}
  \& D.~{Deming}, 173

\bibitem[\protect\citeauthoryear{{Kov{\'a}cs}, {Zucker}, \&
  {Mazeh}}{{Kov{\'a}cs} et~al.}{2002}]{Kovacs_Zucker_Mazeh:aa:2002a}
{Kov{\'a}cs}, G., {Zucker}, S.,  \& {Mazeh}, T. 2002, A\&A, 391, 369

\bibitem[\protect\citeauthoryear{{Lasker} et~al.}{{Lasker}
  et~al.}{1990}]{Lasker_Sturch_McLean:aj:1990a}
{Lasker}, B.~M., {Sturch}, C.~R., {McLean}, B.~J., {Russell}, J.~L., {Jenkner},
  H.,  \& {Shara}, M.~M. 1990, AJ, 99, 2019

\bibitem[\protect\citeauthoryear{{Latham}}{{Latham}}{1992}]{Latham:ASP:1992a}
{Latham}, D.~W. 1992, in ASP Conf. Ser. 32: Complementary Approaches to Double
  and Multiple Star Research, ed. H.~A. {McAlister} \& W.~I. {Hartkopf}, IAU
  Colloq. 135 (San Fransisco: ASP), 110

\bibitem[\protect\citeauthoryear{{Latham}}{{Latham}}{2003}]{Latham:ASP:2003a}
{Latham}, D.~W. 2003, in ASP Conf. Ser. 294: Scientific Frontiers in Research
  on Extrasolar Planets, ed. D.~{Deming} \& S.~{Seager} (San Fransisco: ASP),
  409

\bibitem[\protect\citeauthoryear{{Laughlin} et~al.}{{Laughlin}
  et~al.}{2005}]{Laughlin_Wolf_Vanmunster:apj:2005a}
{Laughlin}, G., {Wolf}, A., {Vanmunster}, T., {Bodenheimer}, P., {Fischer}, D.,
  {Marcy}, G., {Butler}, P.,  \& {Vogt}, S. 2005, ApJ, 621, 1072

\bibitem[\protect\citeauthoryear{{Mandushev} et~al.}{{Mandushev}
  et~al.}{2005}]{Mandushev_Torres_Latham:apj:2005a}
{Mandushev}, G., et~al. 2005, ApJ, 621, 1061

\bibitem[\protect\citeauthoryear{{McCullough} et~al.}{{McCullough}
  et~al.}{2005}]{McCullough_Stys_Valenti:pasp:2005a}
{McCullough}, P.~R., {Stys}, J.~E., {Valenti}, J.~A., {Fleming}, S.~W.,
  {Janes}, K.~A.,  \& {Heasley}, J.~N. 2005, PASP, 117, 783

\bibitem[\protect\citeauthoryear{{McCullough} et~al.}{{McCullough}
  et~al.}{2006}]{McCullough_Stys_Valenti:apj:2006a}
{McCullough}, P.~R., et~al. 2006, ApJ, 648, 1228

\bibitem[\protect\citeauthoryear{{O'Donovan}, {Charbonneau}, \&
  {Kotredes}}{{O'Donovan}
  et~al.}{2004}]{ODonovan_Charbonneau_Kotredes:AIP:2004a}
{O'Donovan}, F.~T., {Charbonneau}, D.,  \& {Kotredes}, L. 2004, in AIP Conf.
  Proc. 713: The Search for Other Worlds, ed. S.~S. {Holt} \& D.~{Deming}, 169

\bibitem[\protect\citeauthoryear{{O'Donovan} et~al.}{{O'Donovan}
  et~al.}{2006a}]{ODonovan_Charbonneau_Mandushev:apjl:2006a}
{O'Donovan}, F.~T., et~al. 2006a, ApJ, 651, L61

\bibitem[\protect\citeauthoryear{{O'Donovan} et~al.}{{O'Donovan}
  et~al.}{2006b}]{ODonovan_Charbonneau_Torres:apj:2006a}
{O'Donovan}, F.~T., et~al. 2006b, ApJ, 644, 1237

\bibitem[\protect\citeauthoryear{{Pepper}, {Gould}, \& {Depoy}}{{Pepper}
  et~al.}{2004}]{Pepper_Gould_Depoy:AIP:2004a}
{Pepper}, J., {Gould}, A.,  \& {Depoy}, D.~L. 2004, in AIP Conf. Proc. 713: The
  Search for Other Worlds, ed. S.~S. {Holt} \& D.~{Deming}, 185

\bibitem[\protect\citeauthoryear{{Pont} et~al.}{{Pont}
  et~al.}{2005}]{Pont_Bouchy_Melo:aa:2005a}
{Pont}, F., {Bouchy}, F., {Melo}, C., {Santos}, N.~C., {Mayor}, M., {Queloz},
  D.,  \& {Udry}, S. 2005, A\&A, 438, 1123

\bibitem[\protect\citeauthoryear{{Pont}, {Zucker}, \& {Queloz}}{{Pont}
  et~al.}{2006}]{Pont_Zucker_Queloz:mnras:2006a}
{Pont}, F., {Zucker}, S.,  \& {Queloz}, D. 2006, MNRAS, 373, 231

\bibitem[\protect\citeauthoryear{{Rauer} et~al.}{{Rauer}
  et~al.}{2004}]{Rauer_Eisloffel_Erikson:pasp:2004a}
{Rauer}, H., {Eisl{\"o}ffel}, J., {Erikson}, A., {Guenther}, E., {Hatzes},
  A.~P., {Michaelis}, H.,  \& {Voss}, H. 2004, PASP, 116, 38

\bibitem[\protect\citeauthoryear{{Seager} \& {Mall{\'e}n-Ornelas}}{{Seager} \&
  {Mall{\'e}n-Ornelas}}{2003}]{Seager_Mallen-Ornelas:apj:2003a}
{Seager}, S.,  \& {Mall{\'e}n-Ornelas}, G. 2003, ApJ, 585, 1038

\bibitem[\protect\citeauthoryear{{Sirko} \& {Paczy{\'n}ski}}{{Sirko} \&
  {Paczy{\'n}ski}}{2003}]{Sirko_Paczynski:apj:2003a}
{Sirko}, E.,  \& {Paczy{\'n}ski}, B. 2003, ApJ, 592, 1217

\bibitem[\protect\citeauthoryear{{Smith} et~al.}{{Smith}
  et~al.}{2006}]{Smith_Collier-Cameron_Christian:mnras:2006a}
{Smith}, A.~M.~S., et~al. 2006, MNRAS, 373, 1151

\bibitem[\protect\citeauthoryear{{Street} et~al.}{{Street}
  et~al.}{2003}]{Street_Pollaco_Fitzsimmons:ASP:2003a}
{Street}, R.~A., et~al. 2003, in ASP Conf. Ser. 294: Scientific Frontiers in
  Research on Extrasolar Planets, ed. D.~{Deming} \& S.~{Seager}, 405

\bibitem[\protect\citeauthoryear{{Szentgyorgyi} et~al.}{{Szentgyorgyi}
  et~al.}{2005}]{Szentgyorgyi_Geary_Latham:AAS:2005a}
{Szentgyorgyi}, A.~H., et~al. 2005, American Astronomical Society Meeting
  Abstracts, 207, 110.10

\bibitem[\protect\citeauthoryear{{Tingley}}{{Tingley}}{2004}]{Tingley:aa:2004a}
{Tingley}, B. 2004, A\&A, 425, 1125

\bibitem[\protect\citeauthoryear{{Tody}}{{Tody}}{1993}]{Tody:1993a}
{Tody}, D. 1993, in ASP Conf. Ser. 52: Astronomical Data Analysis Software and
  Systems II, ed. R.~J. {Hanisch}, R.~J.~V. {Brissenden}, \& J.~{Barnes} (San
  Fransisco: ASP), 173

\bibitem[\protect\citeauthoryear{{Torres} et~al.}{{Torres}
  et~al.}{2004}]{Torres_Konacki_Sasselov:apj:2004b}
{Torres}, G., {Konacki}, M., {Sasselov}, D.~D.,  \& {Jha}, S. 2004, ApJ, 614,
  979

\bibitem[\protect\citeauthoryear{{Udalski} et~al.}{{Udalski}
  et~al.}{2002}]{Udalski_Paczynski_Zebrun:acta:2002a}
{Udalski}, A., et~al. 2002, Acta Astron., 52, 1

\bibitem[\protect\citeauthoryear{{Vidal-Madjar} et~al.}{{Vidal-Madjar}
  et~al.}{2003}]{Vidal-Madjar_Lecavelier-des-Etangs_Desert:nat:2003a}
{Vidal-Madjar}, A., {Lecavelier des Etangs}, A., {D{\'e}sert}, J.-M.,
  {Ballester}, G.~E., {Ferlet}, R., {H{\'e}brard}, G.,  \& {Mayor}, M. 2003,
  Nature, 422, 143

\bibitem[\protect\citeauthoryear{{Young}}{{Young}}{1967}]{Young:aj:1967a}
{Young}, A.~T. 1967, AJ, 72, 747

\bibitem[\protect\citeauthoryear{{Zacharias} et~al.}{{Zacharias}
  et~al.}{2004}]{Zacharias_Urban_Zacharias:aj:2004a}
{Zacharias}, N., {Urban}, S.~E., {Zacharias}, M.~I., {Wycoff}, G.~L., {Hall},
  D.~M., {Monet}, D.~G.,  \& {Rafferty}, T.~J. 2004, AJ, 127, 3043

\end{thebibliography}

\clearpage

\begin{deluxetable}{lccc}
\tablewidth{0pt}
\tablecaption{TrES labels for And0 candidate planets, together with corresponding 2MASS and GSC designations and instrumental $V$ magnitude. \label{tab:names}}

\tablehead{ \colhead{Candidate} & \colhead{2MASS \tablenotemark{a}}  &  \colhead{GSC \tablenotemark{b}} & \colhead{$V$}}

\startdata
\tOne    & \object[2MASS J01083088+4938442]{01083088+4938442} & \object[GSC 03272-00845]{03272--00845} & 11.4 \\
\tTwo    & \object[2MASS J00531053+4717320]{00531053+4717320} & \object[GSC 03266-00642]{03266--00642} & 11.6 \\
\tThree & \object[2MASS J01023745+4808421]{01023745+4808421} & \object[GSC 03267-01450]{03267--01450} & 12.0 \\
\tFour   & \object[2MASS J01180059+4927124]{01180059+4927124} & \object[GSC 03272-00540]{03272--00540} & 12.2 \\
\tFive    & \object[2MASS J00545421+4805505]{00545421+4805505} & \object[GSC 03266-00119]{03266--00119} & 12.7\\
\tSix      & \object[2MASS J00595445+4902030]{00595445+4902030} & \object[GSC 03271-01102]{03271--01102} & 12.7
\enddata
\tablenotetext{a}{Designations from 2MASS Catalog \citep{Cutri_Skrutskie_van-Dyk:2003a}, giving the coordinates of the sources in the form \mbox{hhmmss.ss+ddmmss.s} J2000.}
\tablenotetext{b}{GSC Catalog \citep{Lasker_Sturch_McLean:aj:1990a}.}
\end{deluxetable}

\begin{deluxetable}{lccccccl}
\tablewidth{0pt}
\tablecaption{Transit properties for the six TrES And0 candidates. \label{tab:transits}}

\tablehead{
\colhead{Candidate} & \colhead{SDE} & \colhead{Depth}  & \colhead{Period}  & \colhead{Duration} & \colhead{$N$  \tablenotemark{a}} & \colhead{Telescope(s) \tablenotemark{b}}& \colhead{True Nature} \\
\colhead{} & \colhead{} & \colhead{(mag)} & \colhead{(Days)} & \colhead{(Hours)} & \colhead{} & \colhead{} & \colhead{}
}

\startdata
\tOne & 19.3 & 0.005 & 1.1198 & 1.6  &  7 &  S,T & Eclipsing binary \\
\tTwo & \phn \phn 9.6 \tablenotemark{c} & 0.009 & 4.6619 & 3.4 &  2 & S & A--type star  \\
\tThree & 13.8 &  0.017 & 4.7399 & 3.4 & 5 & S,P,T & Eclipsing binary \\
\tFour & 21.5 & 0.019 & 3.0691 & 2.2 & 3 & S,P & Eclipsing binary \\
\tFive & 12.7 & 0.007 & 2.6540 & 3.2 & 7 & S,P & Blend \\
\tSix & 18.2 & 0.007 & 2.3556 & 3.4 & 7 & S,P,T & Rapidly rotating A--type star
\enddata
\tablenotetext{a}{The number of distinct transits observed in the TrES data set.}
\tablenotetext{b}{TrES telescopes that detected transits of this candidate, where $S$ is Sleuth, $P$ is PSST, and $T$ is STARE.}
\tablenotetext{c}{Here the SDE is based on the Sleuth data set, rather than the TrES combined observations.}
\end{deluxetable}

\begin{deluxetable}{lcccccccc}
\tablewidth{0pt}
\tablecaption{Photometric and spectroscopic properties of the six TrES And0 candidates. \label{tab:catalogs}}

\tablehead{
\colhead{Candidate} & \colhead{$v_{r}$ \tablenotemark{a}} & \colhead{$P(\chi^{2})$ \tablenotemark{b}} & \colhead{$T_{\mathrm{eff}}$ \tablenotemark{c}}  &  \colhead{$\log{g}$ \tablenotemark{c}}  & \colhead{$v\sin{i}$ \tablenotemark{c}}  & \colhead{$\mu$ \tablenotemark{d}} & \colhead{$B_{T}-V_{T}$  \tablenotemark{e}} & \colhead{$J-K$ \tablenotemark{f}}\\
\colhead{} & \colhead{($\mathrm{km\,s^{-1}}$)} & \colhead{} & \colhead{($K$)} & \colhead{} & \colhead{($\mathrm{km\,s^{-1}}$)} & \colhead{($\mathrm{\mathrm{mas\,yr^{-1}}}$)} & \colhead{(mag)} & \colhead{(mag)}
}

\startdata
\tOne & $-28.82\pm\phn0.53$ & 0.180 & 5250 & 3.0 & \phn 3  &  3.0 &  \phs $0.87$  & $0.60$  \\
\tTwo & $-13.05\pm\phn4.06$ & \nodata &  9500 & 4.5 & 55 & 7.7 &  $-0.07$ & $0.01$  \\
\tThree & $\phm{-0}4.47\pm27.87$ & 0.000 & 7000 & 3.5 & 22 &  3.5 & \nodata & $0.24$   \\
\tFour & $-11.31\pm\phn4.73$ & 0.008 & 6250 & 3.5 & 77 & 5.5 & \nodata & $0.14$   \\
\tFive & $-15.41\pm\phn0.31$ & 0.747 & 5500 & 3.5 & \phn 2  & 6.0 & \nodata & $0.66$   \\
\tSix & $-35.26\pm\phn3.43$ & \nodata &  7750 & 3.5 & 88 & 8.2 & \nodata & $0.25$
\enddata
\tablenotetext{a}{The mean radial velocity.}
\tablenotetext{b}{The probability that the observed chi--square should be less than a value $\chi^{2}$, assuming that our model of a star without radial velocity variation is correct.}
\tablenotetext{c}{For a discussion of the errors in these spectroscopic data, see Section~\ref{sec:reject}.}
\tablenotetext{d}{UCAC2 proper motions \citep{Zacharias_Urban_Zacharias:aj:2004a}.}
\tablenotetext{e}{Tycho--2 visible colors \citep{Hog_Fabricius_Makarov:aa:2000b,Hog_Fabricius_Makarov:aa:2000a}.}
\tablenotetext{f}{2MASS infrared colors \citep{Cutri_Skrutskie_van-Dyk:2003a}.}
\end{deluxetable}

\begin{deluxetable}{lccc}
\tablewidth{0pt}
\tablecaption{Time of observation, orbital phase and radial velocity for spectroscopic observations of the six TrES And0 candidates. \label{tab:rv}}

\tablehead{
\colhead{Candidate} & \colhead{Time of Observation} & \colhead{Photometric Orbital Phase} & \colhead{Radial Velocity} \\
\colhead{} & \colhead{HJD} & \colhead{} & \colhead{$\mathrm{km\,s^{-1}}$}
}

\startdata
\tOne   & 2453276.8749 & 0.46  & $-28.16 \pm 0.36$ \\
\nodata & 2453277.8530 & 0.33 & $-29.19 \pm 0.40$ \\
\nodata & 2453278.8236 & 0.20 & $-29.00 \pm 0.32$ \\
\tTwo    & 2453276.8062 & 0.73 & $-13.05 \pm 4.06$ \\
\tThree  & 2453276.8642 & 0.80 & $\phm{-}35.46 \pm 0.74$ \\
\nodata & 2453301.8475 & 0.07 & $-14.14 \pm 0.99$ \\
\nodata & 2453334.7734 & 0.02 & $-\phn3.77 \pm 0.94$ \\
\nodata & 2453548.9711 & 0.21 & $-39.63 \pm 1.12$ \\
\nodata & 2453575.9740 & 0.90 & $\phm{-}24.54 \pm 1.05$ \\
\nodata & 2453576.9683 & 0.11 & $-20.34 \pm 1.16$ \\
\nodata & 2453626.9029 & 0.65 & $\phm{-}23.98 \pm 1.13$ \\
\nodata & 2453627.9043 & 0.86 & $\phm{-}29.37 \pm 0.70$ \\
\nodata & 2453628.8392 & 0.06 & $-\phn9.31 \pm 0.81$ \\
\nodata & 2453629.8518 & 0.27 & $-42.28 \pm 1.12$ \\
\nodata & 2453630.8361 & 0.48 & $-15.29 \pm 1.24$ \\
\nodata & 2453631.8335 & 0.69 & $\phm{-}30.93 \pm 0.86$ \\
\nodata & 2453632.8135 & 0.90 & $\phm{-}24.80 \pm 0.87$ \\
\nodata & 2453633.8047 & 0.10 & $-20.66 \pm 1.02$ \\
\nodata & 2453636.8830 & 0.75 & $\phm{-}37.44 \pm 0.86$ \\
\nodata & 2453779.5864 & 0.84 & $\phm{-}29.64 \pm 1.14$ \\
\tFour   & 2453276.8864 & 0.11 & $-\phn7.55 \pm 1.40$ \\
\nodata & 2453686.7688 & 0.18 & $-14.50 \pm 3.74$ \\
\tFive    & 2453276.8192 & 0.17 & $-15.81 \pm 0.44$ \\
\nodata & 2453277.8381 & 0.55 & $-15.15 \pm 0.41$ \\
\nodata & 2453301.8342 & 0.59 & $-15.09 \pm 0.43$ \\
\nodata & 2453334.7619 & 0.00 & $-15.50 \pm 0.43$ \\ 	
\tSix      & 2453276.8398 & 0.22 & $-35.26 \pm 3.43$ \\
\enddata
\end{deluxetable}

\clearpage

\end{document}